\documentclass[12pt,a4paper]{article}
\usepackage{mathrsfs}
\usepackage{epsfig}
\usepackage{slashed}
\usepackage{amssymb}
\usepackage{amsmath}
\usepackage{amsthm}
\begin{document}
\title{Dark Matter Searches in Jet plus Missing Energy in $\rm \gamma p$ collision at CERN LHC}
\author{Hao Sun\footnote{haosun@mail.ustc.edu.cn \hspace{0.2cm} haosun@dlut.edu.cn} \\
{\small Institute of Theoretical Physics, School of Physics $\&$ Optoelectronic Technology,} \\
{\small Dalian University of Technology, Dalian 116024, P.R.China}
}
\date{}
\maketitle

\begin{abstract}

In this paper, we investigate the $\rm \gamma p$ photoproduction
of jet plus missing energy signal to set limits on the couplings
of the fermionic dark matter to the quarks at the LHC via the main reaction
$\rm pp\rightarrow p\gamma p\rightarrow p \chi\chi j$.
We assume a typical LHC multipurpose forward detectors and work
in a model independent Effective Field Theory framework.
Typically, when we do the background analysis, we also include
their corresponding Single Diffractive (SD) productions.
Our result shows that by requiring a $5\sigma$ ($\rm S/\sqrt{B} \geq 5$)
signal deviation, with an integrated luminosity of $\rm {\cal L} = 200 fb^{-1}$,
the lower bounds of WIMP masses scale can be detected up to $\rm \Lambda$
equal 665.5, 808.9 and 564.0 GeV
for the forward detector acceptances $\xi_1$, $\xi_2$, and $\xi_3$,
respectively, where $0.0015<\xi_1<0.5$, $0.1<\xi_2<0.5$ and $0.0015<\xi_3<0.15$.

 \vspace{0.1cm} \vspace{2.0cm} \noindent
\end{abstract}

\newpage
\section{Introduction}

Cosmological observations imply the existence of dark matter (DM) to be
the dominant component of cosmical matter\cite{DMEvidence1,DMEvidence2}.
Its relic density has been determined precisely by the WMAP experiment
to be $\rm \Omega_\chi h^2 = 0.1109 \pm 0.0056$\cite{RelicDensity} and it is
believed to be non-baryonic, cold, dissipationless and stable on time scales.
In addition to these, their physical properties, like making ups or deep natures,
are still unknown. Revealing the distribution and the nature of dark matter is one of
the most interesting challenges in the fields of both cosmology and particle physics.

Many dark matter candidates have been proposed. Weakly Interacting
Massive Particles (WIMPs) is the most compelling one among them.
Many beyond the Standard Model (BSM) theories, such as Supersymmetry\cite{WIMPsusy},
Warped\cite{WIMPwarped} and Universal\cite{WIMPUED} Extra Dimensions
or Little Higgs Models\cite{WIMPLH}, etc, predict good candidates
for the WIMPs and for the cosmological requirements, i.e., the WIMP
abundance is a natural consequence of the thermal history of the
universe\cite{WIMPcandidates}. Although well motivated, there is still
no experimental evidence to support these theories. It will be difficult
to judge which theory is proper for dark matter. Even the observations
of dark matter itself from future experiments may not provide enough information
to distinguish underlying theories. In this case, a model independent studies of
dark matter phenomenology using Effective Field Theory (EFT) can be particularly
important.

Various experiments are set up to hunt for the particle they make up and to reveal
the nature of dark matter. For instance, through direct detection (DD)
experiments\cite{ExpDMDD1,ExpDMDD2,ExpDMDD3,ExpDMDD4}, we can search for scattering of
dark matter particles from the galactic halo on detector nuclei.
Through indirect detection (ID), we can detect dark matter particles by their annihilation
into high energy Standard Model (SM) particles \cite{DMannihilation}.
In addition, particle production through high energy accelerators
will be another interesting way of dark matter hunting.  
In this case, the dark matter particles are expected to be detected as a missing component, or
manifest as an excess of events showing an imbalance in momentum conservation.
There exist some experimental and theoretical studies include, i.e., the visible radiation of a jet
(quark or gluon)\cite{ExpDMLHCjet1,ExpDMLHCjet2,ExpDMLHCjet3},
a photon\cite{ExpDMLHCphoton1,ExpDMLHCphoton2},
or a W/Z boson decaying into leptons or hadronic jets\cite{ExpDMLHCv1,ExpDMLHCv2,ExpDMLHCv3}
plus the missing "something".

The Large Hadron Collider (LHC) at CERN generates high energetic proton-proton
($\rm pp$) collisions with a luminosity of $\rm {\cal L}=10^{34}cm^{-2}s^{-1}$
and provides the opportunity to study very high energy physics.
In such high energy, most attention is usually paid to the central rapidity
region where the most of the particles are produced and where the most of the
high $\rm p_T$ signal of new physics is expected. Indeed, the CDF collaboration
has already observed such a kind of interesting phenomenon
including the exclusive lepton pairs production \cite{ppllpp1,ppllpp2},
photon-photon production \cite{pprrpp}, dijet production \cite{ppjjpp} and
charmonium ($\rm J/\psi$) meson photoproduction \cite{ppJPHIpp}, etc.
Now, both the ATLAS and the CMS collaborations have programs of forward physics,
which are devoted to studies of high rapidity regions,
with extra updated detectors located in a place nearly 100-400m
close to the interaction point \cite{FDs1,FDs2}.
Technical details of the ATLAS Forward Physics (AFP) projects
can be found, for example, in Refs.\cite{AFP,AFP1}.
The physics program of this new instrumentation covers interesting topics like
elastic scattering, diffraction, low-x QCD,
Central Exclusive Production (CEP), photon-photon ($\gamma\gamma$) and
photo-proton ($\rm \gamma p$) interactions.

Dark matter searching will be an active topic and an important issue
at the LHC\cite{DMsearchLHC1,DMsearchLHC2}.
Until now most works are concentrated on its searching through normal pp collision.
However, it will also be very interesting to see the status of dark matter searching
in photon-photon ($\gamma\gamma$) and photo-proton ($\rm \gamma p$) interactions.
This is mainly due to the reason that photon interactions at the LHC are believed to be simple and
clean from challenged backgrounds. The study of photon interactions at the LHC 
might be a choice of extending the discovery bounds of
dark matter as which that will be shown in our discussion.
In this paper, we focus on the $\rm \gamma p$ photoproduction of jet plus missing energy
signal to set limits on the couplings of the fermionic dark matter to the quarks at the LHC
via the main reaction $\rm pp\rightarrow p\gamma p\rightarrow p \chi\chi j$. We assume
a typical LHC multipurpose forward detectors and work in a model independent
EFT framework. Paper is organized as follow:
we build the calculation framework in Section 2 including a brief introduction
to $\rm \gamma p$ collision and to the WIMP production process we are interested in.
Section 3 is arranged to present the numerical results and background analysis.
Typically, the Single Diffractive (SD) production as background to $\rm \gamma p$ productions
is considered. Finally we summarize our conclusions in the last section.

\section{Calculation Framework}

\subsection{$\rm \gamma p$ Collision at the LHC and Equivalent Photon Approximation}

\begin{figure}[hbtp]
\vspace{-4.4cm}
\hspace*{1cm}
\includegraphics[scale=0.6]{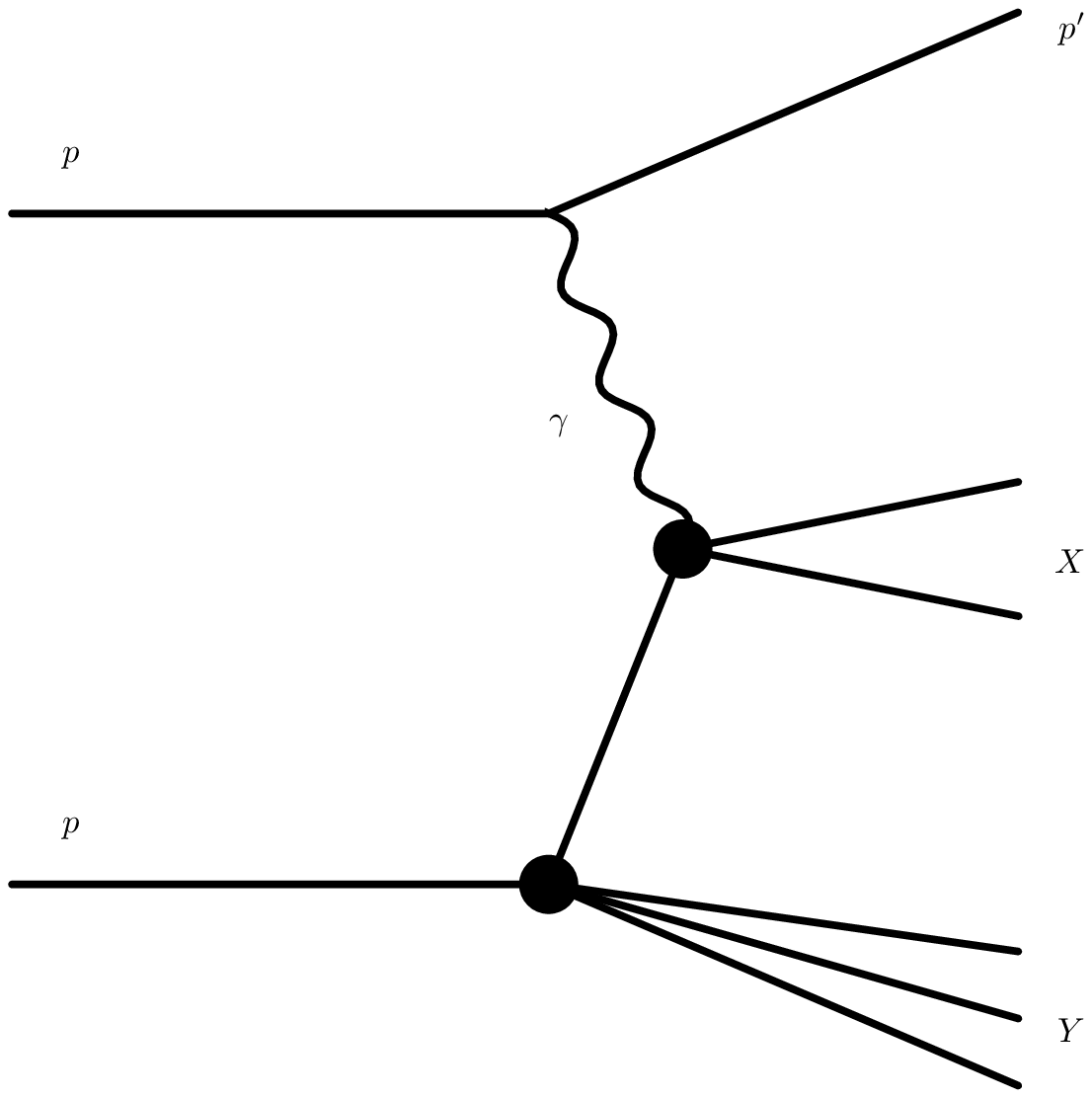}
\vspace{-6.5cm}
\caption{\label{rpexclusive}
A generic diagram for the photoproduction of $\rm pp\rightarrow p\gamma p\rightarrow pXY$
at the CERN LHC.}
\end{figure}

Photoproduction is a class of processes in which one of the two interacting protons is
not destroyed during the collision but survive into the final state with additional particle
(or particles) state(s). Protons of this kind are named intact or forward protons.
A generic diagram for the photoproduction denoted as
$\rm pp\rightarrow p\gamma p\rightarrow pXY$ is presented in Fig.\ref{rpexclusive}.
The kinematics of a forward proton is often described by means of the reduced energy loss $\xi$,
which is also defined as the forward detector acceptance:
\begin{eqnarray}
 \xi=\rm \frac{\Delta E}{E}=\frac{E-E'}{E}
\end{eqnarray}
where E is the initial energy of the beam and $\rm s=4 E^2$ is the square of the centre of mass energy.
$\rm E'$ is the energy after the interaction and $\rm \Delta E$
is the energy that the proton lost in the interaction.

The possibility of adding forward proton detectors to both the ATLAS and the CMS experiments
has received quite some attention, since the forward proton tagging
would provide a very clean environment for new physics searches.
Different from usual $\rm pp$ Deep Inelastic Scattering (DIS),
incoming protons dissociate into partons, jets will be made from the proton remnants
which create some ambiguities and make the new physics signal detection suffer from
incredible backgrounds, $\gamma\gamma$ and $\rm \gamma p$ collisions
can provide more clean environment. In this case, the quasi-real photons
emit with very low virtuality from protons, leave the radiating proton intact,
thus provide an extra experimental handle (forward proton tagging) to help reduce the
backgrounds. If both incoming and emitted protons remain intact, then it provides
the $\gamma\gamma$ collisions, which can be cleaner than the $\rm \gamma p$ collisions.
However, $\rm \gamma p$ collisions have higher energy and effective luminosity
with respect to $\gamma\gamma$ collisions.

Deflected protons and their energy loss will be detected by the forward detectors
with a very large pseudorapidity, while the remaining system will go to central detector.
Photons emitted with small angles by the protons show a spectrum of virtuality $\rm Q^2$
and the energy $\rm E_\gamma$. This is described by the Equivalent Photon Approximation
(EPA)\cite{EPA} which differs from the point-like electron (positron) case by taking care
of the electromagnetic form factors in the equivalent $\gamma$ spectrum and effective
$\gamma$ luminosity:
\begin{equation}\label{EPAformula}
\rm \frac{dN_\gamma}{dE_\gamma dQ^2}=\rm
\frac{\alpha}{\pi}\frac{1}{E_\gamma Q^2}\left[(1-\frac{E_\gamma}{E})(1-\frac{Q^2_{min}}{Q^2})F_E
 + \frac{E^2_\gamma}{2 E^2}F_M \right]
\end{equation}
with
\begin{eqnarray} \nonumber
\rm Q^2_{min}=( \frac{M^2_{inv} E}{E-E_{\gamma}} -M^2_P )\frac{E_{\gamma}}{E},
 ~~~~ F_E= \frac{4 M^2_p G^2_E + Q^2 G^2_M}{4 M^2_p +Q^2}, \\
\rm G^2_E=\frac{G^2_M}{\mu^2_p}=(1+\frac{Q^2}{Q^2_0})^{-4}, ~~~~F_M=G^2_M, ~~~~Q^2_0=0.71 GeV^2 ,
\end{eqnarray}
where $\rm \alpha$ is the fine-structure constant, E is the energy of the incoming proton beam,
which is related to the quasi-real photon energy by $\rm E_\gamma=\xi E$.
$\rm M_p$ is the mass of the proton and $\rm M_{inv}$ is the invariant mass of the final state.
$\rm \mu^2_p$ = 7.78 is the magnetic moment of the proton.
$\rm F_E$ and $\rm F_M$ are functions of the electric and magnetic form factors
given in the dipole approximation.

We denote the photoproduction processes in Fig.\ref{rpexclusive} as
\begin{eqnarray}
\rm pp \rightarrow  p\gamma p \rightarrow p + \gamma + q/\bar{q}/g
\rightarrow p + \underbrace{i + j + k + ...}_X +  Y
\end{eqnarray}
with q = u, d, c, s, b and i, j, k, ... the final state particles.
The hadronic cross section at the LHC can be converted by integrating
$\rm \gamma + q/\bar{q}/g \rightarrow i + j + k +...$ over the photon ($\rm dN(x,Q^2)$),
gluon and quark ($\rm G_{g,q/p}(x_2,\mu_f)$) spectra:
\begin{eqnarray}\label{totcrosssection}  \nonumber
\rm \sigma_{\gamma p}=\rm \sum_{j=q,\bar{q},g} \int^{\sqrt{\xi_{max}}}_{\frac{M_{inv}}{\sqrt{s}}} 2z dz \int^{\xi_{max}}_{Max(z^2,\xi_{min})}
\frac{dx_1}{x_1} \int^{Q^2_{max}}_{Q^2_{min}} \frac{dN_\gamma(x_1)}{dE_\gamma dQ^2} G_{g,q/p}(\frac{z^2}{x_1}, \mu_f) \\
\rm \cdot \int \frac{1}{avgfac} \frac{|{\cal M}_n (\gamma j\rightarrow klm..., \hat s =z^2 s )|^2}{2 \hat s (2 \pi)^{3n-4}} d\Phi_n ,
\end{eqnarray}
where $\rm x_1$ is the ratio between scattered quasi-real photons and incoming proton energy
$\rm x_1 = E_\gamma/E$. $\rm \xi_{min} (\xi_{max})$ are its lower (upper) limits which means
the forward detector acceptance satisfies $\rm \xi_{min}\leq\xi\leq\xi_{max}$.
$\rm x_2$ is the momentum fraction of the proton momentum carried by the gluon (quark).
The quantity $\rm \hat s = z^2 s$ is the effective center-of-mass system (c.m.s.) energy with
$\rm z^2=x_1 x_2$. $\rm s=4 E^2$ mentioned above and $\rm M_{inv}$ is the total mass
of the related final states. $\rm 2z/x_1$ is the Jacobian determinant
when transform the differentials from $\rm dx_1dx_2$ into $\rm dx_1dz$. $\rm G_{g,q/p}(x,\mu_f)$
represent the gluon (quark) parton density functions, $\rm \mu_f$ is the factorization scale.
$\rm f=\frac{dN}{dE_\gamma dQ^2}$
is the $\rm Q^2$ dependent relative luminosity spectrum present in Eq.(\ref{EPAformula}).
$\rm Q^2_{max}=2 GeV^2$ is the maximum virtuality.
$\rm \frac{1}{avgfac}$ is the times of the spin-average factor, the color-average factor
and the identical particle factor. $\rm |{\cal M}_n|^2$ presents the squared n-particle matrix element
and divided by the flux factor $\rm [2 \hat s (2 \pi)^{3n-4}]$. The n-body phase space differential
$\rm d\Phi_n$ and its integral $\rm \Phi_n$ depend only on $\rm \hat s$
and particle masses $\rm m_i$ due to Lorentz invariance:
\begin{eqnarray} \nonumber
\rm \Phi_n(\hat s, m_1, m_2, ..., m_n) &=&\rm  \int d\Phi_n(\hat s, m_1, m_2,..., m_n) \\
&=&\rm  \int \delta^4((p_i+p_j)-\sum^{n}_{k=1}p_k) \prod^{n}_{k=1}d^4 p_k \delta(p^2_k-m^2_k) \Theta (p^0_{k})
\end{eqnarray}
with i and j denoting the incident particles and k running over all outgoing particles.

A brief review of experimental prospects for studying
photon induced interactions are summarized in Ref.\cite{HEPhotonIntatLHC}.
Many other related phenomenological studies are summarized
here: standard model productions\cite{SMWH,rbWtVtb1,rbWtVtb2},
supersymmetry\cite{SUSYprrp1,SUSYprrp2},
extra dimensions\cite{EDpllp1,EDprrp2,EDprrp3,EDrqrq4},
unparticle physics\cite{unparticle}, top triangle moose model\cite{TTMrbtp},
gauge boson self-interactions
\cite{anoWWr1,anoWWr2,anoVVV,anoWWrr,anoWWr3,anoZZZ,anoZZrr,anoZZrZrr,anoWWrrZZrrrp,anoVVVV},
neutrino electromagnetic properties\cite{electromagnetic1,electromagnetic2,electromagnetic3},
the top quark physics\cite{Anomaloustqr,Anomaloustqr1,AnomalousWtb,rbWtVtb1,rbWtVtb2}
and triplet Higgs production\cite{TripletH}, etc.

\subsection{WIMPs production via $\rm pp\rightarrow p\gamma p\rightarrow p \chi\chi j$}

In our study, we assume that the dark matter candidate is the only new particle
which is singlet under the SM local symmetries, and all SM particles are singlets
under the dark-sector symmetries. The interaction between dark matter and
the SM particles are presumably effected by the exchange of some heavy mediators
whose nature we do not need to specify, but only assume that they are much heavier
than the typical scales. We specialize to the case of a spin-$1/2$, Dirac fermion
and describe the interaction between dark matter and SM quarks accurately
in terms of an Effective Field Theory\cite{EFTDM}.
There consists of the SM Lagrangian plus kinetic terms for the dark matter $\chi$
and a set of effective four fermion interactions between $\chi$ and the quarks q = u, d, s, c, b, t:
\begin{eqnarray}\label{LEFTDM}
\rm {\cal L}_{\chi\chi qq} = \sum_q  \left[ \frac{1}{\Lambda^2_{D5}}
   \bar{q}\gamma^\mu q \bar{\chi} \gamma_\mu \chi
 + \frac{1}{\Lambda^2_{D8}} \bar{q}\gamma^\mu \gamma_5 q \bar{\chi} \gamma_\mu \gamma_5 \chi  \right]
\end{eqnarray}
where $\chi$ refers to the dark matter particle.
The characterizing parameters $\rm \Lambda_{i}$ are the scales of the effective interactions
where $\rm \Lambda_i=\frac{M}{\sqrt{g_\chi g_q}}$ with M is the mass of the mediator,
$\rm g_\chi$ is the coupling to dark matter particle
and $\rm g_q$ is its coupling to SM quarks.
The first term in Eq.(\ref{LEFTDM}) corresponds to the spin-independent scattering with vector coupling
while the second term corresponds to the spin-dependent axial-vector coupling.
Here we will consider interactions that yield a spin-independent scattering cross section
and simplify our consideration to the vector coupling case.
Similarly, our discussion can be easily extended to the other cases.

\begin{figure}[hbtp]
\vspace{-6cm}
\hspace*{-3cm}
\centering
\includegraphics[scale=0.7]{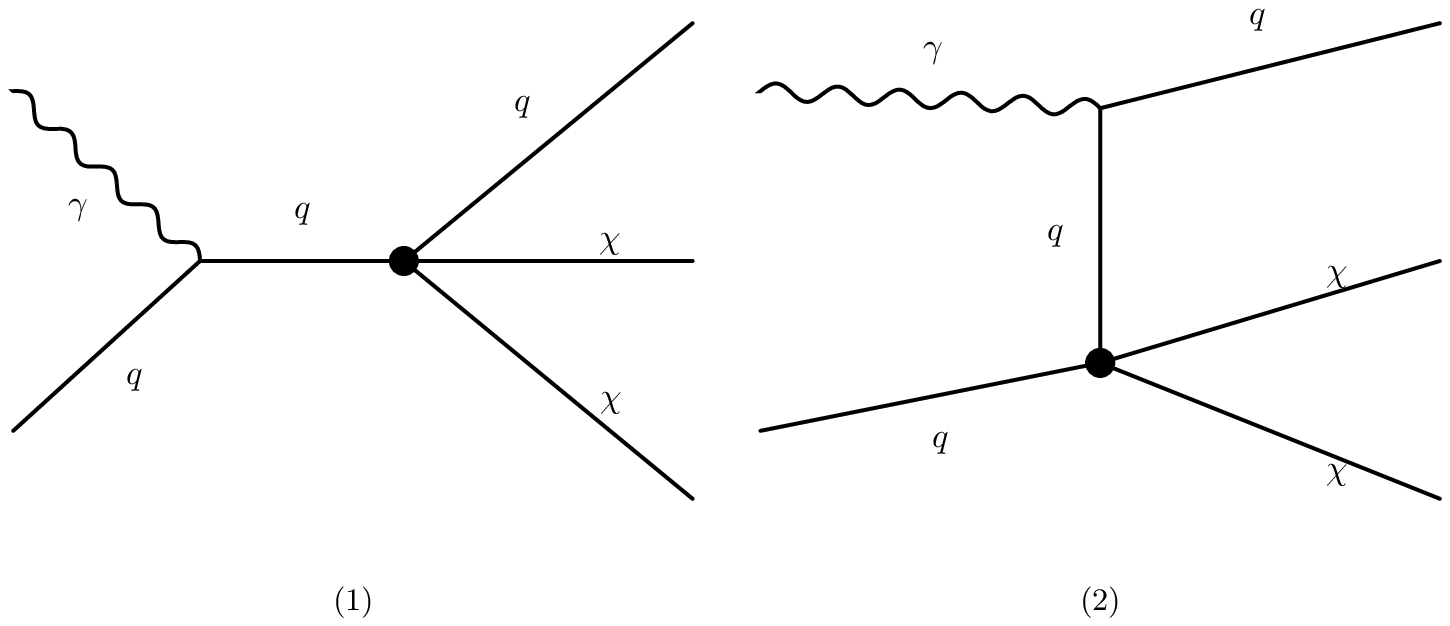}
\vspace{-9cm}
\caption{\label{dm} Parton level Feynman diagrams for
$\rm pp\rightarrow p\gamma p\rightarrow p \chi\chi j$
include the s-channel contribution (1) and the t-channel contribution (2).
Here q=u, d, c, s, b, $\bar{u}$, $\bar{d}$, $\bar{c}$, $\bar{s}$ and $\bar{b}$.
Black blobs represent the $\rm \chi\chi q q$ couplings parameterized by Eq.(\ref{LEFTDM}).
}
\end{figure}

In the effective field theory, WIMP-SM interactions lead to the direct production of
WIMPs via the process $\rm pp\rightarrow \chi\bar\chi$ through the normal pp collision mode at the colliders.
In order to contain a visible trace, one can study WIMPs produced together with
hard jet(s), $\gamma$(s) or heavy gauge boson (W, Z) and provide a trigger within hard scatterings.
Different from normal pp collision, no direct production of $\rm pp\rightarrow \chi\bar\chi$
appear at the $\rm \gamma p$ collider. The simplest production is
$\rm pp\rightarrow p\gamma p\rightarrow p+\chi\bar\chi j$.
Here jet can not be $\gamma$ or gluon but only light quarks. The Feynman diagrams can be found in
Fig.\ref{dm}. We implement the lagrangian into FeynArts and use the Feynman diagram approach
to provide the amplitudes. By inserting the amplitudes into Eq.(\ref{totcrosssection})
we can obtain the total photon produced cross section for
$\rm pp\rightarrow p\gamma p\rightarrow p \chi\chi j$ at the LHC.

\section{Numerical Results}

\subsection{Input Parameters and Kinematic Cuts}

We take the input parameters as $\rm M_p=0.938272046\ GeV$,
$\rm \alpha_{ew}(m^2_Z)^{-1}|_{\overline{MS}}=127.918$,
$\rm m_Z=91.1876 GeV$, $\rm m_W=80.385\ GeV$,
$\rm \Gamma_W=2.085\ GeV$, $\rm \Gamma_Z=2.4952\ GeV$\cite{2012PDG}
and we have $\rm sin^2\theta_W=1-(m_W/m_Z)^2=0.222897$.
Light fermion masses are neglected through the whole calculation.
The colliding energy in the proton-proton c.m.s. system is assumed to be
$\rm \sqrt{s}=14\ TeV$ at the future LHC with luminosity taken to be a running parameter.
The factorization scale is chosen to be $\rm \mu_f=M_Z$.
The strong coupling constant $\rm \alpha_s$ = 0.118.
We use FeynArts, FormCalc and LoopTools (FFL)\cite{FeynArts,FormCalc,LoopTools}
packages to generate the amplitudes and perform the numerical calculations
for both the signal and the background, with in-house modification that needed.
We adopt CT10\cite{CT10} PDF for the parton distributions during calculation
and BASES\cite{BASES} to do the phase space integration.
Based on the forward proton detectors to be installed by the CMS-TOTEM and
the ATLAS collaborations we choose the detected acceptances to be
\begin{itemize}
 \item CMS-TOTEM forward detectors with $0.0015<\xi_1<0.5$
 \item CMS-TOTEM forward detectors with $0.1<\xi_2<0.5$
 \item AFP-ATLAS forward detectors with $0.0015<\xi_3<0.15$
\end{itemize}
which we simply refer to as $\xi_1$, $\xi_2$ and $\xi_3$, respectively.
Such forward detected acceptances will force the proton intact
together with the $\rm p_T$ cuts of the related proton.
In practically the detector (i.e., AFP) is independent of $\rm p_T$
but much sensitive mainly to the $\xi$ value.
These acceptances with the following kinematical cuts
will implicit the selected events to have an intact proton and to have been tagged by the forward detectors.
During calculation we use $\xi_1$ in default unless otherwise stated.
Note that the gap survival probability ($\rm |S|^2$)\cite{SurvivalP}
in forward productions has some process dependence.
With early measurements of rapidity gap processes at the LHC,
they will provide valuable information on $\rm |S|^2$.
We expect that $\rm |S|^2_{\gamma\gamma}>|S|^2_{\gamma p}>|S|^2_{SD}$.
In our phenomenon study the survival probability for
$\rm \gamma p$ collision production is chosen to be $\rm |S|^2_{\gamma p}=0.7$,
while that for the Single Diffractive (see following) production 
is chosen to be $\rm |S|^2_{SD}=0.12$.

In all the results present in our work, we impose a cut of pseudorapidity $|\eta| < 2.5$
for the final state particles since central detectors of the ATLAS and CMS
have a pseudorapidity $|\eta|$ coverage 2.5.
The general acceptance cuts for the events are:
\begin{eqnarray} \label{cuts}
&&\rm p_T^j \geq 25GeV, \slashed{E}_T^{miss} > 30 GeV, |\eta^j|<2.5.
\end{eqnarray}
In order to reduce the backgrounds containing missing energy
from processes where, for example, W bosons are produced,
we remove such events for which a charged lepton has transverse momenta
$\rm p_T^\ell \geq 25GeV$ and is further separated
from all jets by $\rm \Delta R (\ell j)>0.4$\cite{DMboundpp},
where $\rm \Delta R = \sqrt{\Delta \Phi^2 + \Delta \eta^2}$ is the separation
in the rapidity-azimuth plane, $\rm p_T^{j,\ell}$ are the transverse momentum
of jets and leptons and $\rm \slashed E_T^{miss}$ the missing transverse momentum.
We stress here that cuts in Eq.({\ref{cuts}}) are the basic ones and might change
later in our following discussion.

\subsection{Background Analysis}

As can be seen, the studied topology of our signal therefore give rise to the
$\rm jet+\slashed E_T$ signature characterized by one jet and
a missing transverse momentum ($\rm \slashed E_T$) from the undetected dark matter pair production.
The dominant SM backgrounds consist of the following types:
\begin{itemize}
 \item
$\rm \gamma p$ collision of the SM production of Z+jet where Z decays into a pair of neutrinos:
\begin{eqnarray}
\rm pp\rightarrow p\gamma p \rightarrow p+ (Z\rightarrow \nu \bar\nu) j + Y
\end{eqnarray}
 \item
$\rm \gamma p$ collision of the SM W+jet associated production with the W decays into a neutrino
and a charged lepton falls outside of the acceptance range of the detector:
\begin{eqnarray}
\rm pp\rightarrow p\gamma p \rightarrow p+ (W\rightarrow \ell^\pm \nu) j + Y
\end{eqnarray}
 \item
Single Diffractive (SD) production of Z+jet as main background to $\rm \gamma p$ collision
where Z decays into a pair of neutrinos.
 \item
Single Diffractive (SD) production of W+jet associated production with the W decays into a neutrino
and the charged lepton falls outside of the acceptance range of the detector.
\end{itemize}

\begin{figure}[hbtp]
\vspace{-6cm}
\hspace*{-3cm}
\centering
\includegraphics[scale=0.7]{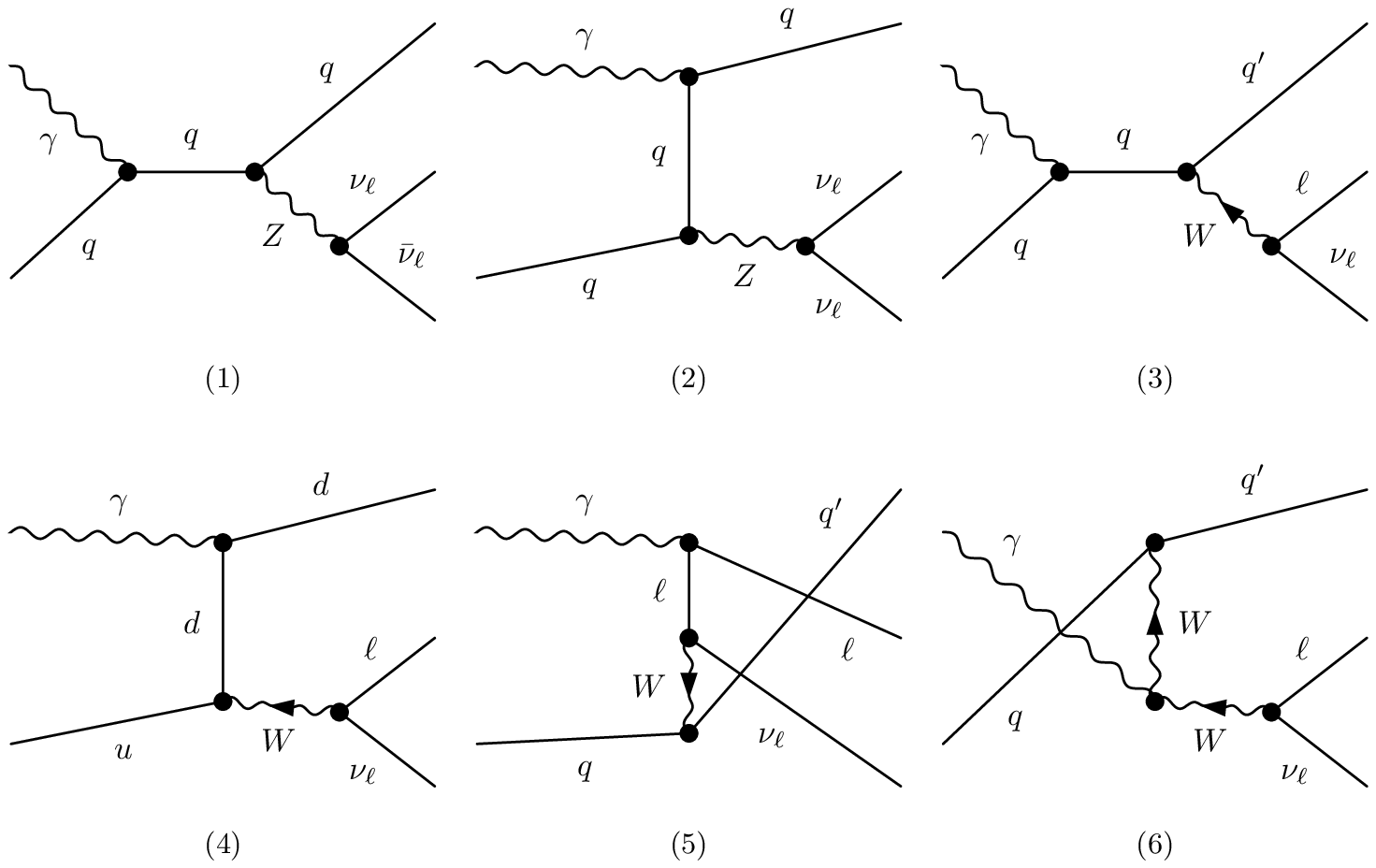}
\vspace{-8cm}
\caption{\label{rqzw}
Feynman diagrams for $\rm \gamma p$ collision of the SM production of Z+jet
where Z decays into a pair of neutrinos and W+jet associated production
with the W decays into a neutrino and a charged lepton falls outside the
acceptance range of the detector..
}
\end{figure}

The first two in the list belong to the SM electroweak processes
with some of their Feynman diagrams presented in Fig.\ref{rqzw}.
The cross section can be obtained by Eq.(\ref{totcrosssection}).
All the three generations of neutrino $\nu_{e,\mu,\tau}$ are taken into account.
The CKM matrix $V_{qq'}$ between different generation of quarks are omitted and
assumed to be unit within the same generation.

Now lets turn to the last two backgrounds listed ahead whose contributions
come from the Single Diffractive productions.
SD production is indeed a kind of main background in photoproductions at the $\rm \gamma p$ collider.
However, their contributions are usually omitted and not often well considered in
some other phenomenological studies. In our paper, we take some efforts to include them
and try to find out how important of their contributions will stand in background analysis
at the $\rm \gamma p$ collider.

In order to provide a clear image to describe the SD production, in a logical clear way,
we start from the description of Double Pomeron Exchange (DPE) production
as the main background of $\gamma\gamma$ collision.
In DPE, two colorless objects are emitted from both protons.
Their partonic components are resolved and create a heavy mass object
X ($\rm pp\rightarrow p\gamma\gamma p\rightarrow p X p$)
in the central detector in the Pomeron-Pomeron interaction.
The event is characterized by two rapidity gaps between the central object and the protons.
Through the exchange of two Pomerons, a dijet system, a diphoton system, a WW and
ZZ pair, or a Drell-Yan pair can be created, for instance.
The DPE production cross section can be obtained within the factorized Ingelman-Schlein\cite{ISmodel} model.
In the Ingelman-Schlein model, one assumes that the Pomeron has a well-defined partonic structure,
and that the hard process takes place in the Pomeron-proton/proton-Pomeron (single diffraction)
or Pomeron-Pomeron (central diffraction) processes.
These processes are described in terms of the proton "diffractive" parton distribution functions (DPDFs).
Similar to the parton distribution functions,
the DPDFs are also obtained from Deep Inelastic Scattering experiments\cite{H1DPDF1,H1DPDF2}.
The difference is that they contain a dependence on some additional variables 
which are used to describe the proton kinematics:
the relative energy loss $\xi$ and four-momentum transfer t, which we introduce as
\begin{eqnarray}\nonumber
\rm g^D(x,\mu^2) &=&\rm \int dx_{\textbf{P}} d\beta \delta(x-x_{\textbf{P}}\beta)
g_{\textbf{P}}(\beta, \mu^2) f_{\textbf{P}}(x_{\textbf{P}})    \\
&=&\rm  \int^1_x \frac{dx_{\textbf{P}}}{x_{\textbf{P}}} f_{\textbf{P}}(x_{\textbf{P}})
g_{\textbf{P}}(\frac{x}{x_{\textbf{P}}},\mu^2)
\end{eqnarray}
where $\rm g^D$ denotes either the quark or the gluon distributions with D refers to "diffractive".
$\rm f_{\textbf{P}}(x_{\textbf{P}})$ is the flux of Pomerons and expressed as
\begin{eqnarray}
\rm f_{\textbf{P}}(x_{\textbf{P}})=\int^{t_{max}}_{t_{min}} f(x_{\textbf{P}},t)  dt
\end{eqnarray}
with $\rm t_{min}$ and $\rm t_{max}$ being the kinematic boundaries.
$\rm g_{\textbf{P}}(\beta,\mu^2)$ is the partonic structure of Pomeron.
$\rm x_{\textbf{P}}$ here is the fraction of the proton momentum carried by the Pomeron corresponding
to the relative energy loss $\xi$. And $\beta$ is the fraction of the Pomeron momentum carried by the struck parton.
Both Pomeron flux factors $\rm f(x_{\textbf{P}},t)$ as well as quark, gluon distributions
in the Pomeron $\rm g_{\textbf{P}} (\beta, \mu^2)$ were taken
from the H1 collaboration analysis of diffractive structure function
at HERA\cite{H1DPDF1,H1DPDF2}.
Therefore, the final forward detector acceptance $\xi$-dependent
convolution integral for the DPE production is given by
\begin{eqnarray}\label{Eq.DPE} \nonumber
\rm \sigma_{DPE} &=& \rm \sum_{ij=q\bar{q},gg,qg,\bar{q}g}  \int^{\xi_{max}}_{\frac{M_{inv}}{\sqrt{s}}} 2z dz \int^{Min(\xi_{max},
\frac{z^2}{\xi_{min}})}_{Max(\frac{z^2}{\xi_{max}},\xi_{min})} \frac{dx_1}{x_1}
\int^1_{x_1} \frac{dx_{\textbf{P}_i}}{x_{\textbf{P}_i}} f_{\textbf{P}_i}(x_{\textbf{P}_i})
g_{\textbf{P}_i}(\frac{x_1}{x_{\textbf{P}_i}},\mu^2) \\
&& \rm  \int^1_{\frac{z^2}{x_1}} \frac{dx_{\textbf{P}_j}}{x_{\textbf{P}_j}} f_{\textbf{P}_j}(x_{\textbf{P}_j})
g_{\textbf{P}_j}(\frac{z^2}{x_1 x_{\textbf{P}_j}},\mu^2)
 \int \frac{1}{avgfac} \frac{|{\cal M}_n (ij\rightarrow klm..., \hat s =z^2 s )|^2}{2 \hat s (2 \pi)^{3n-4}} d\Phi_n.
\end{eqnarray}
Logically it will now easy to turn to SD production.
A Pomeron-proton or proton-Pomeron interaction can be used to describe
it and the total SD production cross section can be directly obtained by
\begin{eqnarray}\label{Eq.SD}  \nonumber
\rm \sigma_{SD}=\rm \sum_{ij=q\bar{q},gg,qg,\bar{q}g}
\int^{\sqrt{\xi_{max}}}_{\frac{M_{inv}}{\sqrt{s}}} 2z dz \int^{\xi_{max}}_{Max(z^2,\xi_{min})}
\frac{dx_1}{x_1} \int^1_{x_1} \frac{dx_{\textbf{P}_i}}{x_{\textbf{P}_i}} f_{\textbf{P}_i}(x_{\textbf{P}_i})
g_{\textbf{P}_i}(\frac{x_1}{x_{\textbf{P}_i}},\mu^2) \\
\rm G_{g,q/p}(\frac{z^2}{x_1}, \mu_f)
\rm \int \frac{1}{avgfac} \frac{|{\cal M}_n (i j\rightarrow klm..., \hat s =z^2 s )|^2}{2 \hat s (2 \pi)^{3n-4}} d\Phi_n.
\end{eqnarray}
Diffraction usually dominates for $\xi<0.05$ which means we can replace
$\rm Min(\xi_{max}, \frac{z^2}{\xi_{min}})$ in Eq.(\ref{Eq.DPE})
or $\rm \xi_{max}$ in Eq.(\ref{Eq.SD}) by 0.05 directly. We check
this and find the difference is quite small as expected. Due to this reason,
we can find that SD background are dominant for $0.0015<\xi_1<0.5$ and
$0.0015<\xi_3<0.15$ while strongly suppressed for $0.1<\xi_2<0.5$.

\begin{figure}[hbtp]
\vspace{-4cm}
\hspace*{-3cm}
\centering
\includegraphics[scale=0.8]{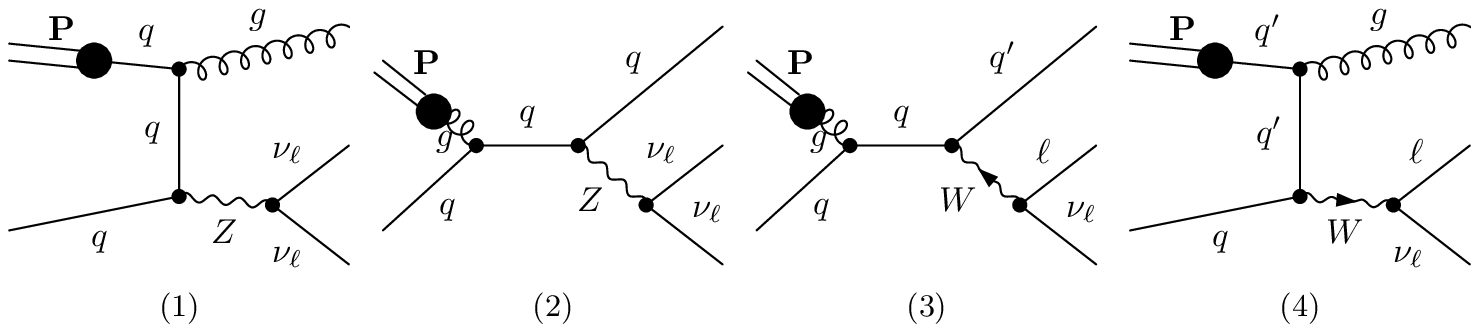}
\vspace{-16cm}
\caption{\label{sdzw} Some Feynman diagrams for Single Diffractive production of Z+jet
where Z decays into a pair of neutrinos and W+jet associated production
with the W decays into a neutrino and a charged lepton falls outside the
acceptance range of the detector.
}
\end{figure}

Feynman diagrams for Single Diffractive production of Z+jet
with Z decays into a pair of neutrinos and for that of W+jet associated production
with the W decays into a neutrino and a charged lepton (where the lepton falls outside the
acceptance range of the detector), are presented in Fig.\ref{sdzw}.
The double solid lines refer to the Pomeron ($\textbf{P}$) that come from the forward proton and going
with a parton (quarks or gluon) emission from inside. The other initial parton comes from the
dissociate proton, described in terms of the normal proton PDF.
The other diagrams with the inverse of the Pomeron-proton, which means replaced by proton-Pomeron,
are similar thus not shown in the figures. Final cross section of the SD productoin
can than be obtain by Eq.(\ref{Eq.SD}). 
We add these productions into FormCalc and perform our calculation also 
with the help of the FFL packages.

\subsection{Data and Signal Boundary}

\begin{figure}[hbtp]
\vspace{1cm}
\centering
\includegraphics[scale=0.6]{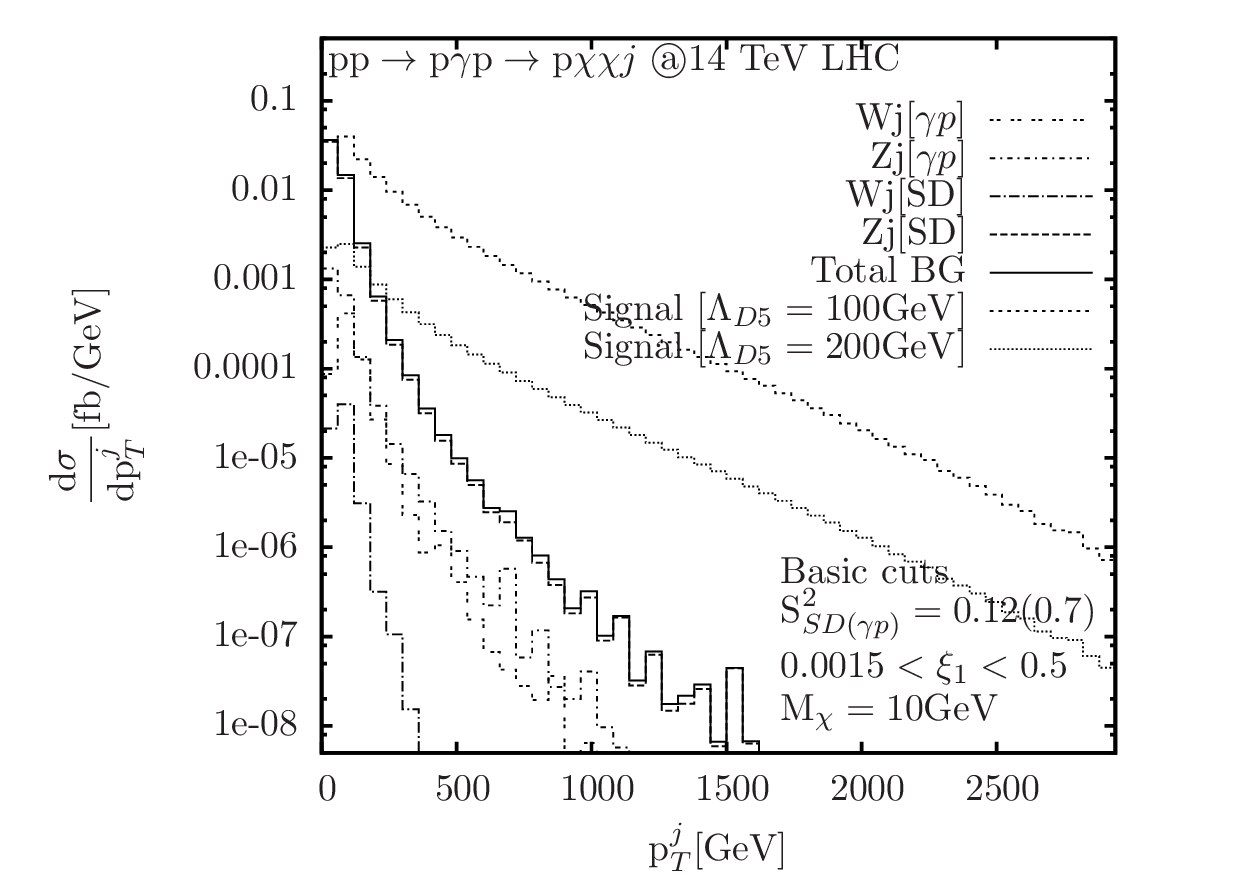}
\includegraphics[scale=0.6]{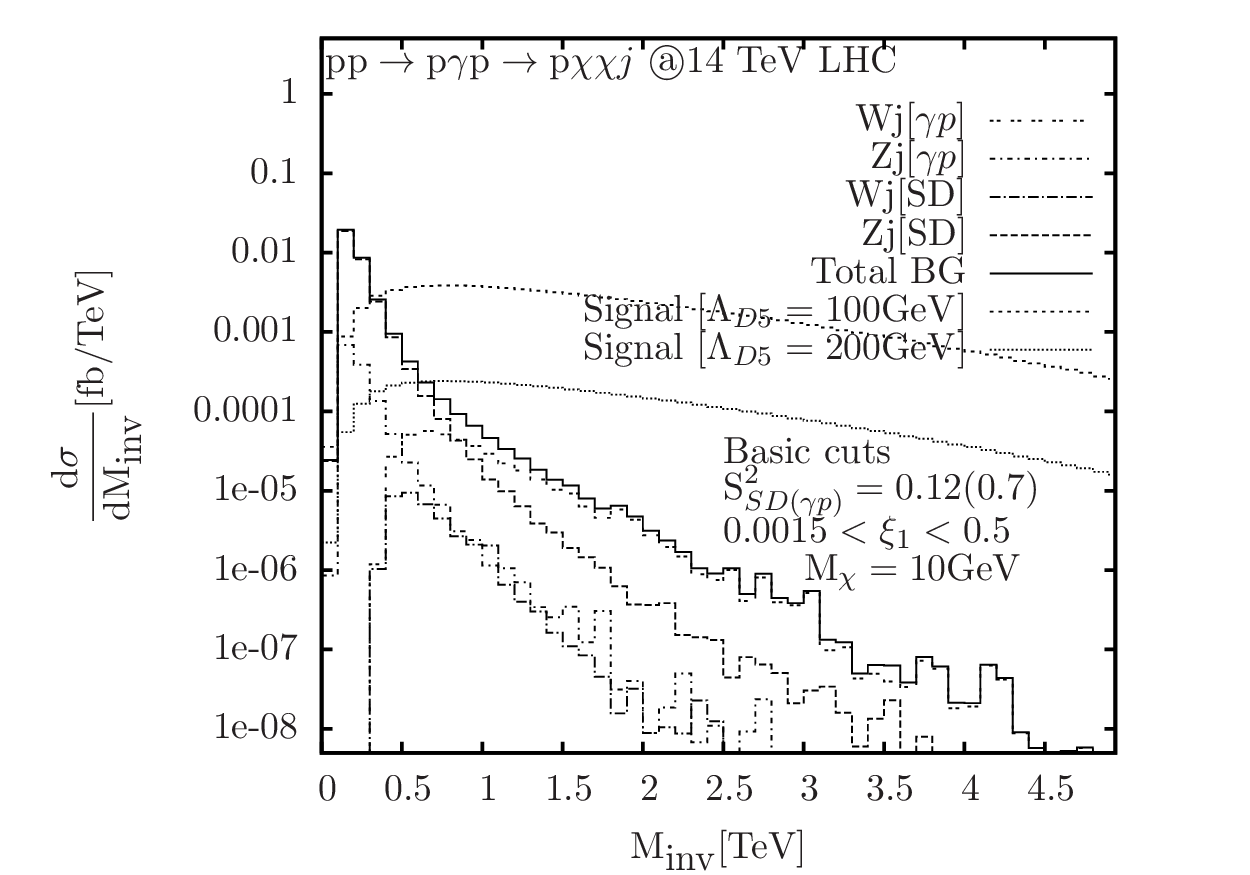}
\includegraphics[scale=0.6]{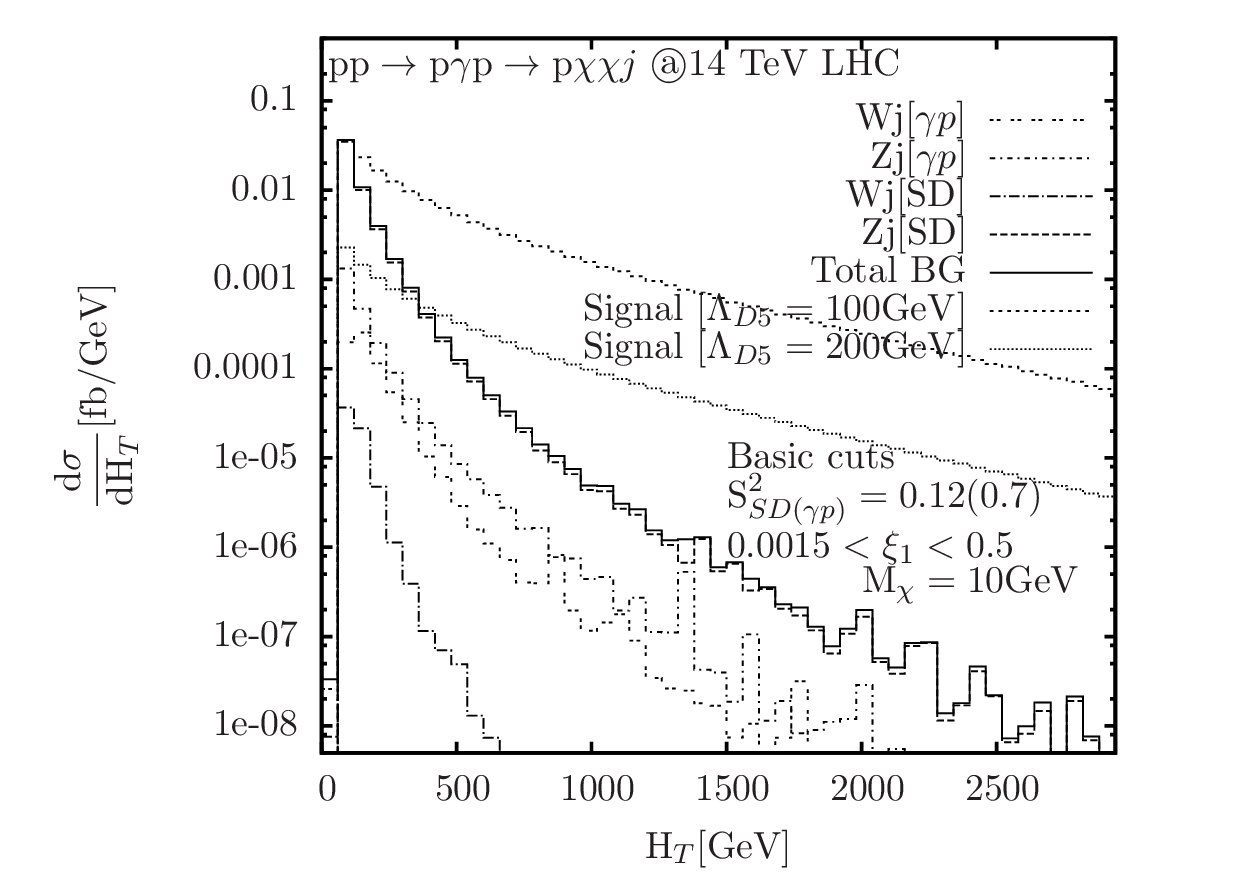}
\caption{
\label{distri}
The signal and background differential cross sections of the transverse momentum
of the jet ($\rm p^{jet}_T$), the invariant mass of the produced System ($\rm M_{inv}$) and
the $\rm H_T$ distributions for $\rm pp\rightarrow p\gamma p\rightarrow p \chi\chi j$.
The Basic kinematic cuts and survival probability factor are taken into account.
The forward detector acceptance is chosen to be $0.0015<\xi_1<0.5$.
Here $\rm M_\chi=10GeV$ and $\rm \Lambda_{D5}=100$ and 200 GeV.}
\end{figure}

In Fig.\ref{distri} we present the differential cross sections of the transverse momentum
of the jet ($\rm p^{jet}_T$), the invariant mass of the produced System ($\rm M_{inv}$) and
the $\rm H_T$ distributions where $\rm H_T=\sum|p^{jet}_T|+\slashed E_T$.
For the signal, we have chosen a WIMP mass of 10 GeV and the effective scale $\rm \Lambda_{D5}=100$ and 200 GeV.
The basic kinematic cuts in Eq.(\ref{cuts})
and the survival probability factor $\rm {|S|^2}_{\gamma p}=0.7$ and
$\rm {|S|^2}_{SD}=0.12$ are taken into account.
The forward detector acceptance is chosen to be $0.0015<\xi_1<0.5$.
Different sub-contributions of the backgrounds
including $\rm \gamma p$ electroweak productions and SD productions are presented.
Their total contributions are depicted by the solid lines.
Compare these background contributions we can see for $0.0015<\xi_1<0.5$,
the dominant background indeed come from SD Z+jet production.
The smallest one is the photon produced W+jet production.
The $\rm \gamma p$ Z+jet production and SD W+jet production are comparable
and not deviate far from each other.
No matter which one, all the backgrounds are dominant over the low $\rm p^{jet}_T$ region
and reduced rapidly through the high $\rm p^{jet}_T$ region.
The same behavior can be found for distributions of $\rm M_{inv}$ and $\rm H_T$.
The dashed and dotted lines present the signal with $\rm \Lambda_{D5}=100\ GeV$ and
$\rm \Lambda_{D5}=200\ GeV$, respectively. All the $\rm p^{jet}_T$, $\rm M_{inv}$ and $\rm H_T$
distributions are enhanced by the dark matter effects at larger or higher regions.
This potential of distributions clearly provide very simple
and effective way to separate the signal from the backgrounds.
This is easily and primarily realized by adopting large $\rm p^{jet}_T$ cuts.
Just to remind here that in Fig.\ref{distri} we choose
the basic kinematic cuts where $\rm p^{jet}_T>25\ GeV$.

In order to see how the cross sections for signal and background depend on the final
transverse momentum jet ($\rm p_T^{jet}$) cuts,
we present this dependence in Tab.\ref{SSx1}
with typical cuts of $\rm p^{jet}_T>25\ GeV$, $\rm p^{jet}_T>200\ GeV$,
$\rm p^{jet}_T>350\ GeV$ and $\rm p^{jet}_T>500\ GeV$, respectively.
We can see that larger $\rm p_T^{jet}$ cut can reduce the background
cross section essentially while make the signal reduce slightly.
As can be found that the larger $\rm p_T^{jet}$ cut can improve the quantity of
the signal over background ratio (S/B),
which implies a result that is more robust against systematic uncertainties.
We also calculate the statistical significance (SS) for the signal and background
on different values of $\rm p_T^{jet}$ cuts in the same table for $\rm \Lambda_{D5}=300\ GeV$
with the following formula\cite{SSformula}:
\begin{eqnarray}\label{SSformula}
\rm SS=\sqrt{2 [(S+B) log(1+\frac{S}{B}) - S] }
\end{eqnarray}
where S and B are the numbers of signal and background events, respectively.
$\cal L$ presents the luminosity of future 14 TeV LHC
where we take the value of 5 and 20 $\rm fb^{-1}$.
We can see the statistical significance of the dark matter effect is large and might be detected.
In the following calculation, we apply $\rm p_T^{jet}>350\ GeV$ for $0.0015<\xi_1<0.5$.
The same table but different values is presented for $0.1<\xi_1<0.5$ in Tab.\ref{SSx2}.
In this case, we apply $\rm p_T^{jet}>200\ GeV$.
From our former studies\cite{Anomaloustqr1} we know that
the production behavior for $0.0015<\xi_1<0.15$ is very close to that of $0.0015<\xi_1<0.5$.
This has also been calculated and confirmed.
Results are not shown here to avoid duplicate. Similarly here we adopt $\rm p_T^{jet}>350\ GeV$.

\begin{table}
\begin{center}
\begin{tabular}{c c c c c }
\hline\hline
\multicolumn{5}{c}{$\sigma$ and SS dependence on $\rm p_T^{jet}$ for $0.0015<\xi_1<0.5$} \\
\hline
$\sigma$[pb] & $\rm p_T^j>25 GeV$ &  $\rm p_T^j>200 GeV$  & $\rm p_T^j>350 GeV$  & $\rm p_T^j>500 GeV$ \\
$\rm \sigma_{BG}$ & 3.2725 & 4.0933 $\times 10^{-2}$ & 5.1713 $\times 10^{-3}$ & 1.1613$\times 10^{-3}$ \\
$\rm \sigma_{{\Lambda_{D5}}=100}$ & 9.1687 & 2.9393 &1.4921 & 8.3580 $\times 10^{-1}$ \\
$\rm \sigma_{{\Lambda_{D5}}=300}$ & 1.1323 $\times 10^{-1}$ &3.6288 $\times 10^{-2}$ & 1.8421 $\times 10^{-2}$
                       &1.0318 $\times 10^{-2}$ \\
\hline
$\rm SS^{{\cal L}=5}_{{\Lambda_{D5}}=300}$ & 4.40079 & 11.2813 & 13.1863 & 12.6422 \\
$\rm SS^{{\cal L}=20}_{{\Lambda_{D5}}=300}$ & 8.80158 & 22.5625 & 26.3726 & 25.2844 \\
\hline\hline
\end{tabular}
\end{center}
\caption{\label{SSx1}
The dependence of total signal[$\rm {\Lambda_{D5}}=100\ GeV$ and 300 GeV], background cross sections and
statistical significance [$\rm {\Lambda_{D5}}=300\ GeV$] for
$\rm pp\rightarrow p\gamma p\rightarrow p \chi\chi j$ on typical values of $\rm p_T^{jet}$ cuts.
Here $\rm M_{\chi}=10\ GeV$ and unit of the cross section is in pb.
The other kinematic cuts and survival probability factor are taken into account.
Forward detector acceptance is chosen to be $0.0015<\xi_1<0.5$.
}
\end{table}

\begin{table}
\begin{center}
\begin{tabular}{c c c c c }
\hline\hline
\multicolumn{5}{c}{$\sigma$ and SS dependence on $\rm p_T^{jet}$ for $0.1<\xi_2<0.5$} \\
\hline
$\sigma$[pb] & $\rm p_T^j>25 GeV$ &  $\rm p_T^j>200 GeV$  & $\rm p_T^j>350 GeV$  & $\rm p_T^j>500 GeV$ \\
$\rm \sigma_{BG}$ & 2.8022$\times 10^{-2}$ & 1.6099$\times 10^{-3}$ & 2.4527$\times 10^{-4}$
        & 7.2497$\times 10^{-5}$ \\
$\rm \sigma_{{\Lambda_{D5}}=100}$ & 4.3063 &1.9020 & 1.1068 & 6.8305 $\times 10^{-1}$ \\
$\rm \sigma_{{\Lambda_{D5}}=300}$ & 5.3164 $\times 10^{-2}$ &2.3481 $\times 10^{-2}$&1.3664$\times 10^{-2}$
                       & 8.4327$\times 10^{-3}$ \\
\hline
$\rm SS^{{\cal L}=5}_{{\Lambda_{D5}}=300}$ & 18.2204  & 21.3139 & 20.6161 & 17.9147 \\
$\rm SS^{{\cal L}=20}_{{\Lambda_{D5}}=300}$ & 36.4408  & 42.6277 & 41.2322 & 35.8294 \\
\hline\hline
\end{tabular}
\end{center}
\caption{\label{SSx2}
The dependence of total signal[$\rm {\Lambda_{D5}}=100\ GeV$ and 300 GeV], background cross sections and
Statistical significance [$\rm {\Lambda_{D5}}=300\ GeV$] for
$\rm pp\rightarrow p\gamma p\rightarrow p \chi\chi j$ on typical values of $\rm p_T^{jet}$ cuts.
Here $\rm M_{\chi}=10\ GeV$ and unit of the cross section is in pb.
The other kinematic cuts and survival probability factor are taken into account.
Forward detector acceptance is chosen to be $0.1<\xi_1<0.5$.
}
\end{table}

\begin{figure}[hbtp]
\centering
\includegraphics[scale=0.7]{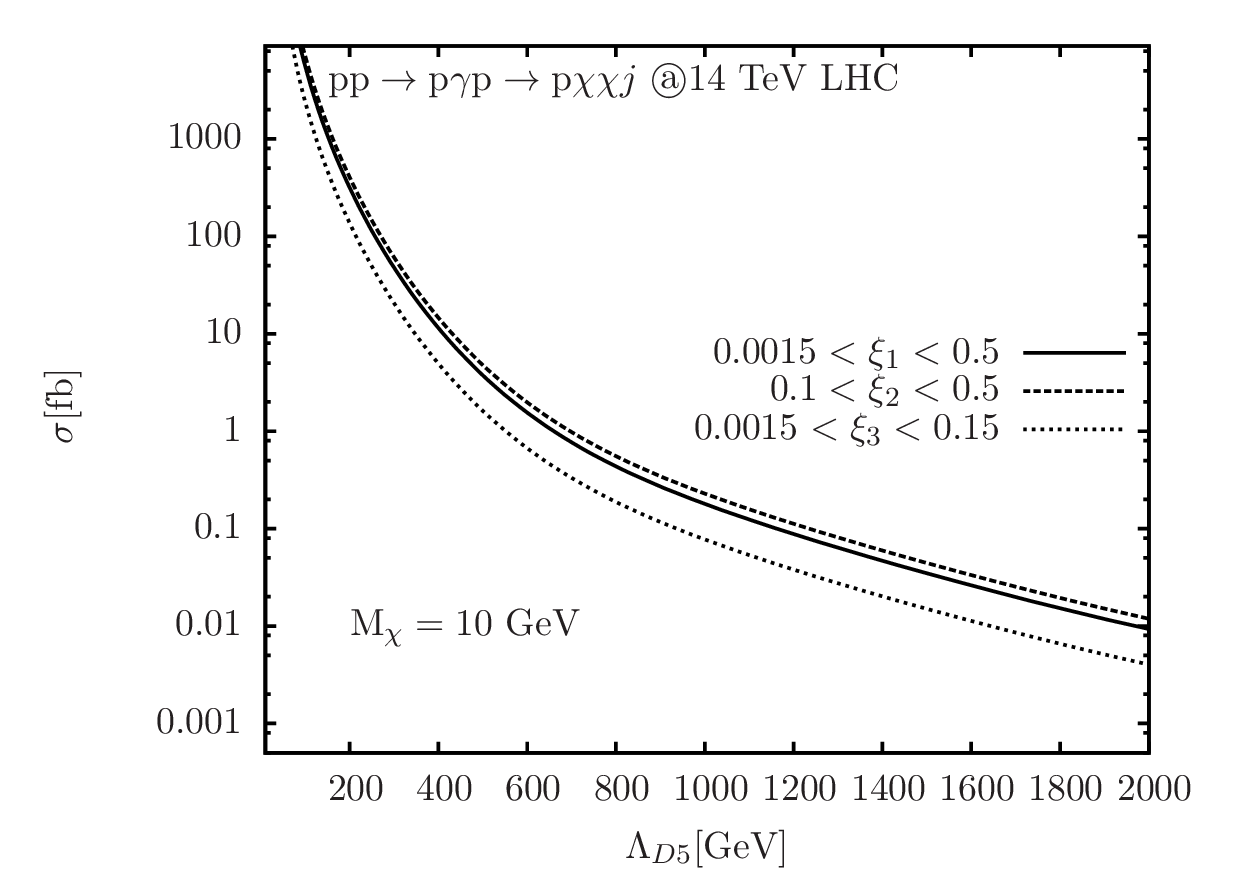}
\caption{\label{mxcrosssection}
The total cross sections of signal process
$\rm pp\rightarrow p\gamma p\rightarrow p \chi\chi j$
as functions of the dark matter effective scale $\rm \Lambda_{D5}$ and three forward detector
acceptance regions: $0.0015<\xi_1<0.5$, $0.1<\xi_2<0.5$ and $0.0015<\xi_3<0.15$.
Notice here we choose $\rm p_T^{jet}>350, 200, 300\ GeV$ for $\xi_1$,
$\xi_2$ and $\xi_3$, respectively. The dark matter particle mass is $\rm M_\chi=10\ GeV$. }
\end{figure}

In Fig.\ref{mxcrosssection}, we present the signal cross sections of
$\rm pp\rightarrow p\gamma p\rightarrow p \chi\chi j$
as functions of the dark matter scale $\rm \Lambda_{D5}$
and three forward detector acceptance: $0.0015<\xi_1<0.5$, $0.1<\xi_2<0.5$ and $0.0015<\xi_3<0.15$.
Solid, dashed and dotted lines correspond to $\xi_1$, $\xi_2$ and $\xi_3$, respectively.
Their cross sections dropped rapidly as $\rm \Lambda_{D5}$ become larger.
For results of $\rm \Lambda_{D5}$ and $\rm \Lambda_{D8}$,
their contributions are indeed very close to each other.
That's why we only focus our study on $\Lambda_{D5}$.
Compare different acceptance regions, though they have almost the same features,
$\xi_1$ and $\xi_3$ do not differ much from each other while both
of them are much larger than cross section of $\xi_2$.
Just to remind that here we choose $\rm p_T^{jet}>350, 200, 350\ GeV$ for $\xi_1$,
$\xi_2$ and $\xi_3$, respectively.
That's why cross section of $\xi_1$ is smaller than that of $\xi_2$,
otherwise if with the same value of $\rm p_T^{jet}$ it will inverse.
The SM backgrounds for the main reactions are
$\sigma_{B}=5.1713~{\rm fb}$ for $0.0015<\xi_1<0.5$,
$\sigma_{B}=1.6099~{\rm fb}$ for $0.1<\xi_2<0.5$ and
$\sigma_{B}=3.8062~{\rm fb}$ for $0.0015<\xi_3<0.15$.
These numbers will be used in our following data analysis.
From this point, we see deviation of the dark matter production
from the backgrounds are obvious which might detectable from future experiments.

\begin{figure}[hbtp]
\centering
\includegraphics[scale=0.7]{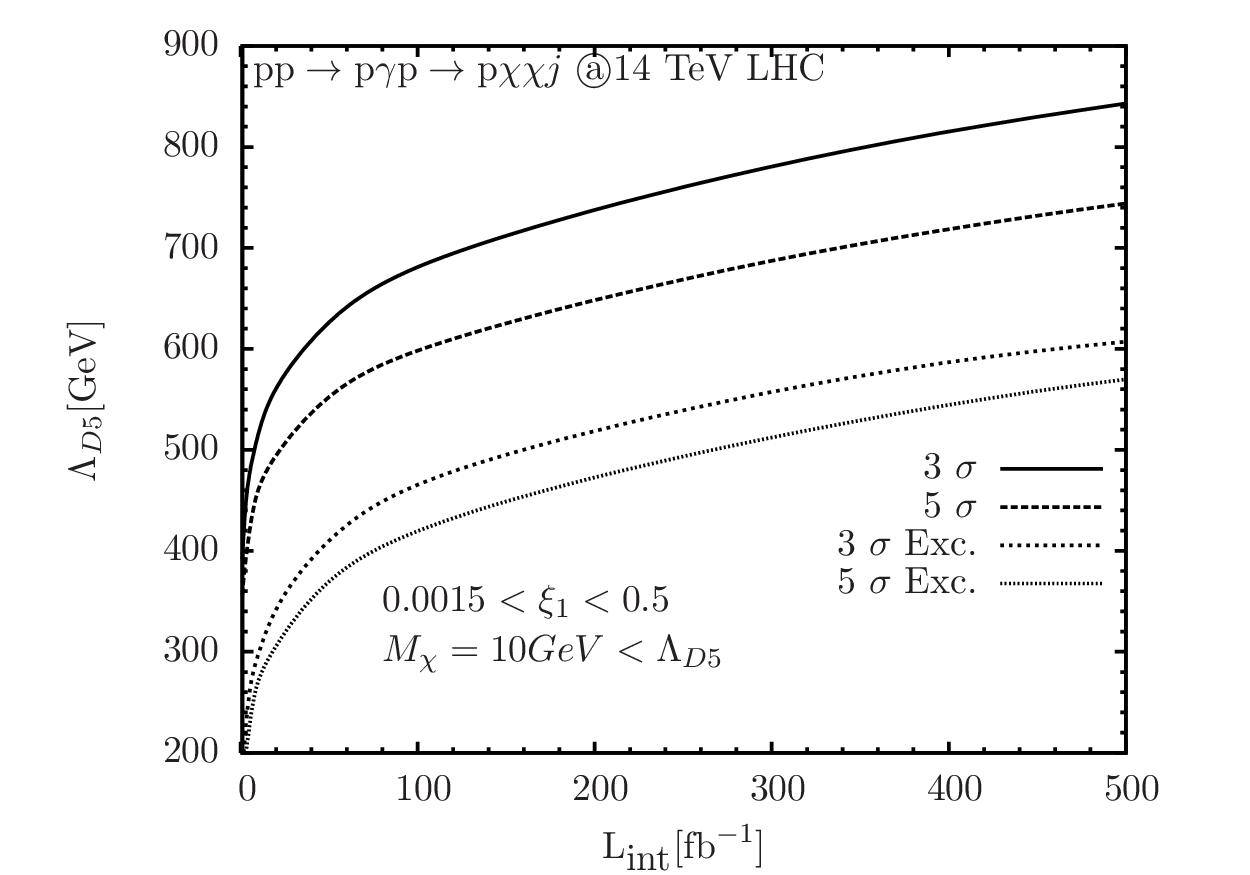}
\includegraphics[scale=0.7]{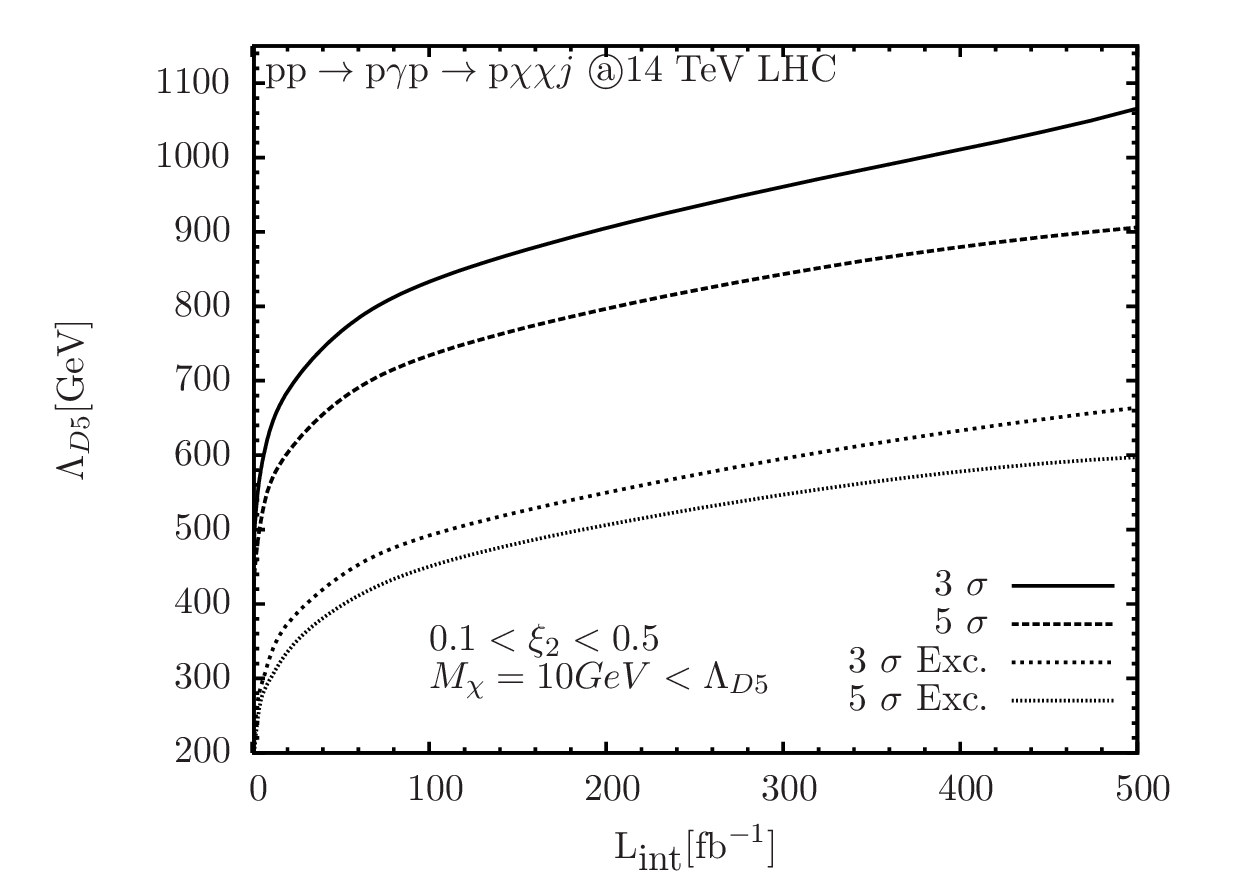}
\includegraphics[scale=0.7]{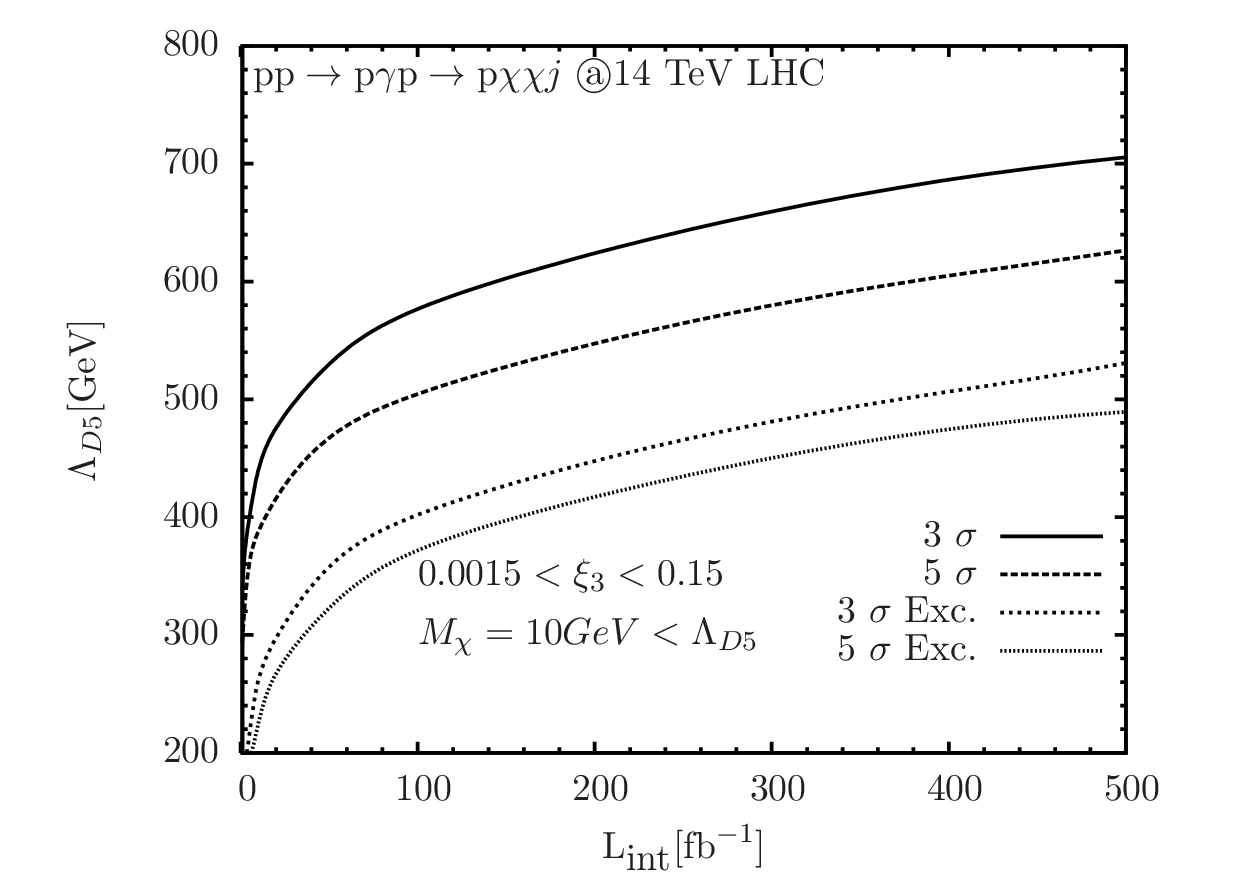}
\caption{\label{mxbound}
$3\sigma$ and $5\sigma$ bounds for the dark matter
effective scale in the $\rm {\cal L}-\Lambda_{D5}$ parameter space,
and $3\sigma$ and $5\sigma$ limit of exclusive boundary
where effective field theory may not valid.
The dark matter particle mass is $\rm M_\chi=10\ GeV$.}
\end{figure}

Now we turn to analyze the potential of the $\rm \gamma p$ collision
at the high energy LHC to probe the dark matter signal through
photoproduction of $\rm \chi\chi+jet$.
We define the $3\sigma$ ($5\sigma$) discovery significance as
\begin{eqnarray}
\rm \frac{|\sigma_{S+B}-\sigma_{B}|}{\sqrt{\sigma_B}} \sqrt{\cal L} > 3(5)
\end{eqnarray}
where ${\cal L}$ is the integrated luminosity of the $\rm \gamma p$ option of the LHC.
In the $\rm {\cal L}-\Lambda_{D5}$ parameter space, we present the plots for the $3\sigma$
and $5\sigma$ deviations of the signal from the total backgrounds.
We display our results for the forward detector acceptance chosen
to be $0.0015<\xi_1<0.5$, $0.1<\xi_2<0.5$ and $0.0015<\xi_3<0.15$
in Fig.\ref{mxbound}, the first, the second and the third panel, respectively.
The center of mass energies of the LHC collider is $\sqrt{s}$ = 14 TeV.
Concerning the criteria 5$\sigma$ deviation of signal from background,
with an integrated luminosity $\rm {\cal L} = 100 fb^{-1}$,
$\rm pp\rightarrow p\gamma p\rightarrow p \chi\chi j$
production through jet photoproduction can probe the dark matter effects
up to $\rm \Lambda_{D5}$ equal 602.7, 748.1 and 510.2 GeV for $\xi_1$, $\xi_2$ and $\xi_3$.
With an integrated luminosity $\rm {\cal L} = 200 fb^{-1}$
one can probe the dark matter effects up to the mass scale
$\rm \Lambda_{D5}$ equal 665.5, 808.9, 564.0 GeV for $\xi_1$, $\xi_2$ and $\xi_3$, respectively.
Here we choose $\rm M_\chi=10\ GeV$, much less than the scale $\Lambda$.
We use solid and dashed lines to present the $3\sigma$ and $5\sigma$ boundary, respectively.
Compare these three forward detector acceptances
$\xi_2$ is the most efficient one to detect the dark matter effects.
This is mainly because $\xi > 0.5$ belong to the region where
single diffractive productions are strongly suppressed,
leaving electroweak photoproductions the main background which are
indeed quite small with our kinematic cuts, thus lead to a better signal over background ratio.
Assuming a 3$\sigma$ deviation of signal from background, for $\xi_2$,
$\rm \gamma p$ collision of $\rm \chi\chi+jet$ can probe $\Lambda$ up to $\rm \sim 1 TeV$,
a factor of 2.5 above the best current bounds from LEP-2\cite{DMboundLEP}.
Our bounds are comparable with the phenomenological study at a 250 GeV ILC,
however, worse than the polarized beams case\cite{DMboundee}.
Similar studies at the normal pp collider can be found, i.e., in Ref.\cite{DMboundpp}
where the boundary of dark matter mass scale can be probe up to 450 GeV at 14 TeV LHC
with the a luminosity of 200 $\rm fb^{-1}$.
Further more, CMS and ATLAS have both carried out the
cut-and-count based monojet\cite{ExpDMLHCjet2,ExpDMLHCjet3}
and monophoton\cite{ExpDMLHCphoton1,ExpDMLHCphoton2} analyses (using $\rm \sim 5 fb^{-1}$ of data).
Typically, for the searches through monophoton channel,
in the case of D5 spin-independent operator,
values of scale below 585 GeV and 156 GeV are excluded at $90\%$ C.L. for
$\rm M_\chi$ equal to 1 GeV and 1.3 TeV, respectively.
Our results show that with the same value of luminosity,
the bound obtained is comparable with that obtained by the forward detector acceptance $\xi_1, \xi_3$,
while smaller than that obtained by $\xi_2$.
It should be noted that the contact operator approximation
may not be valid if the energy scale of the operator
is smaller than the energy scale of the collider process.
The bounds we have already obtained may only be somewhat heuristic\cite{IanShoemaker}.
This means that an experiment can exclude such a range of $1/\hat{s}>1/\Lambda^2$.
In our case the limit is relevant for collider dark matter searches,
thus $\sqrt{\hat{s}}$ here is the invariant mass of the invisibles.
In Fig.\ref{mxbound} we also present the $3\sigma$ and $5\sigma$
limit boundary (with dotted and short dashed curves, respectively)
of the range that should be excluded where the effective field theory may not valid.

In Fig.\ref{mxbound}, we also present the bounds obtained when the luminosity becomes
larger than $\rm 200 fb^{-1}$, see, up to $\rm 500 fb^{-1}$.
However, we should mention here that as the luminosity become larger,
identify the signal under the high pileup running conditions will be challenge:
the hadronic background of multiple p-p interactions will be so large that any $\rm \gamma p$
process will be completely swamped. This can be a drawback of $\rm \gamma p$ productions.
Even though, purpose of our study does not mean to argue that $\rm \gamma p$ production
are better or not to probe dark matter than the other colliders.
Instead, the main purpose is to show the
potential of $\rm \gamma p$ productions on the dark matter searching.
By applying the dark matter searching on the $\rm \gamma p$ production
with forward proton tagging we can have at least a general image of the
dark matter searching through $\rm \gamma p$ collision
after the running of LHC but before the building of future $\rm e^-e^+$ colliders.

Moreover, the production channel we consider here ($\rm \gamma q\rightarrow \chi\chi j$) is
related by crossing symmetry to the hard process considered in monophoton dark
matter searches ($\rm q\bar{q}\rightarrow \chi\chi \gamma$).
The advantage of $\rm \gamma q\rightarrow \chi\chi j$ production
is the reduction in background associated with tagging the intact proton.
While the advantage of the monophoton production
would be that: a photon in the final state may be cleaner than a jet,
and one is more likely to get a $\rm q/\bar{q}$ from the initial proton than a photon.
It will also be interesting to study a monophoton search in which the initial q or $\rm \bar{q}$
arises from pomerons which would have a potentially larger production cross-section,
and a cleaner final state.
A similar study has been presented in Ref.\cite{pppZp} where they
concentrate on the Standard Model Z boson production. It can also be extended
to the study of new physics, i.e., dark matter searches.
A detailed discussion on kinds of these productions might be beyond the
scope of this work, we address here the possibility and interesting to study them,
and will work on this in a parallel paper.

\section{Summary}

Dark matter searches at the high energy colliders have received a lot of attention.
The study of WIMPs in a model independent Effective Field Theory (EFT) approach is of
particularly interesting. Even if a process can be traced back to a definite set
of operators, it is rarely the case that a particular collider signature
can be traced back to a unique process. For this reason many different, complementary
measurements on different collider modes and production channels are usually required to uncover
the underlying new physics.
LHC provides opportunities to open new fields of the study on the $\gamma\gamma$
and $\rm \gamma p$ collisions with very high energy but very low backgrounds
thanks to the forward proton projects.
Attention can be paid now to physics in the central rapidity region
where the most of the high $\rm p_T$ signal of new physics is expected.

In this paper, we investigate the $\rm \gamma p$ photoproduction
of jet plus missing energy signal to set limits on the couplings
of the fermionic dark matter to the quarks at the LHC via the main reaction
$\rm pp\rightarrow p\gamma p\rightarrow p \chi\chi j$.
We assume a typical LHC multipurpose forward detectors and work
in a model independent EFT framework.
Typically, when we do the background analysis, we also include
their corresponding Single Diffractive (SD) productions.
Our result shows that by requiring a $5\sigma$ ($\rm S/\sqrt{B} \geq 5$)
signal deviation, with an integrated luminosity of $\rm {\cal L} = 200 fb^{-1}$,
the lower bounds of the dark matter mass scale can be probed up to
$\rm \Lambda_{D5}$ equal 665.5, 808.9, 564.0 GeV for the forward detector acceptances
$\xi_1 (\xi_2, \xi_3)$, respectively,
where $0.0015<\xi_1<0.5$, $0.1<\xi_2<0.5$ and $0.0015<\xi_3<0.15$.
Assuming a 3$\sigma$ deviation of signal from background, for $\xi_2$,
$\rm \gamma p$ collision of $\rm \chi\chi+jet$ can probe $\Lambda$ up to $\rm \sim 1 TeV$,
a factor of 2.5 above the best current bounds from LEP-2 can comparable with
similar studies at the normal pp colliders.
By applying the dark matter searching on the $\rm \gamma p$ production with forward proton tagging
we present a general image of the dark matter searching at the $\rm \gamma p$ collision
after the running of LHC but before the building of future $\rm e^-e^+$ colliders.

\section*{Acknowledgments} \hspace{5mm}
The author thanks Fawzi Boudjema for his kindness to provide invaluable comments,
thanks Mao Song for useful discussions.
Project supported by the National Natural Science Foundation of China (No. 11205070),
Shandong Province Natural Science Foundation (No. ZR2012AQ017)
and by the Fundamental Research Funds for the Central Universities (No. DUT13RC(3)30).

\end{document}